# Transforming Platform-Independent to Platform-Specific Component and Connector Software Architecture Models

Jan Oliver Ringert[2], Bernhard Rumpe[1] and Andreas Wortmann[1]
[1] Software Engineering, RWTH Aachen University, http://www.se-rwth.de/
[2] School of Computer Science, Tel Aviv University, http://www.cs.tau.ac.il/

*Abstract*—Combining component & connector architecture description languages with component behavior modeling languages enables modeling great parts of software architectures platform-independently. Nontrivial systems typically contain components with programming language behavior descriptions to interface with APIs. These components tie the complete software architecture to a specific platform and thus hamper reuse. Previous work on software architecture reuse with multiple platforms either requires platform-specific handcrafting or the effort of explicit platform models. We present an automated approach to transform platform-independent, logical software architectures into architectures with platform-specific components. This approach introduces abstract components to the platform-independent architecture and refines these with components specific to the target platform prior to code generation. Consequently, a single logical software architecture model can be reused with multiple target platforms, which increases architecture maturity and reduces the maintenance effort of multiple similar software architectures.

## I. Introduction

Component & connector (C&C) architecture description languages (ADLs) [1] combine component-based software engineering with model-driven engineering (MDE) to describe complex software systems as interacting components. Describing component behavior with modeling languages enables to model great parts of software architectures platform-independently. Complex systems, however, require components that interface with APIs to access operating system functions or hardware drivers. Describing the behavior of such components with abstract modeling languages is hardly feasible. Instead, their behavior usually is defined in terms of general purpose programming languages (GPLs). Using GPL components in an architecture ties it to these GPLs and the interfaced APIs. This hampers reuse with different platforms.

Current approaches to generative MDE with C&C ADLs either do not take multi-platform reuse into account [2]–[6] or require explicit platform models [7]–[9]. The former requires duplicating the software architecture and changing the affected components manually, which introduces maintenance and evolution efforts as the duplicated architectures need to be fixed and progressed. The latter introduces complex notions to describe models of the target platform and the mapping of components to it. This introduces efforts in definition, maintenance, and evolution of platform models.

We present an approach to transform platform-independent, logical software architectures into platform-specific architectures of the same modeling language prior to code generation. With this, single logical software architectures can be reused with similar target platforms easily. This approach exploits the black-box nature of components by introducing *abstract components*. These provide stable interfaces to the software architecture, but omit behavior implementations to act as extension points for platform-specific components. Hence, generation of executable systems from such architectures is impossible. Prior to code generation, the abstract components are thus bound to compatible platform-specific components and the software architecture is transformed accordingly. The resulting platform-specific architecture is a well-formed, type-safe model available to further analyses and existing code generators can transform it into executable systems.

Our approach is implemented with the MontiArc-Automaton [10]–[12] C&C ADL and introduces a modeling language to describe *bindings* of software architecture models as well as different *library types*. It builds upon previous work presented in [13] and presents the following improvements:

- architectures and bindings are transformed to type-safe architectures before code generation instead of relying on special annotations of the abstract syntax,
- binding to platform-specific components may add platform-specific parameters,
- code generators need not be aware of replacement of implementations as we transform the architecture prior to code generation (generators process plain architectures),
- code libraries and library models are replaced with implementation libraries, which contribute platform-specific components instead.

This contribution presents the new approach and explains the model transformation to translate platform-independent architectures into platform-specific architectures. To this end, Sect. II describes the required preliminaries of MontiArc-Automaton before Sect. III motivates multi-platform generative MDE by example. Afterwards, Sect. IV introduces the new notions of bindings and libraries. Sect. V relies on these to describe the transformation from platform-independent to platform-specific software architecture models. Finally, Sect. VI discusses related work, including differences to our previous approach, and Sect. VIII concludes.

## II. The MontiArcAutomaton C&C Architecture Modeling Framework

MontiArcAutomaton is a modeling framework for C&C software architectures with application-specific component behavior languages that features a powerful code generation framework. The modeling language comprises a C&C

ADL [11], embeds a component behavior modeling language based on I/O$^\omega$ automata [11], and uses UML/P class diagrams [14] to model data types. It describes logically distributed software architectures in which components perform computations and connectors regulate communication. Components are black-boxes with stable interfaces of typed, directed ports and are either atomic or composed. Atomic components contain a component behavior description, either as a model of an embedded language [12], or as a reference to a GPL artifact. Composed components contain a hierarchy of subcomponents and their behavior emerges from subcomponent interaction. Components do not reveal whether they are composed or atomic or whether they feature a behavior model.

MontiArcAutomaton distinguishes component types from their instantiation and supports component configuration parameters. Component types define the interface and subcomponents of all their instances. Configuration parameters resemble constructors from object-oriented programming and serve component instantiation. Their arguments are passed by the containing component type. Component types may extend other component types and inherit their interfaces and component configuration parameters. Inheriting types may introduce new ports and configuration parameters. Atomic component types may extend composed component types and vice versa. Each atomic component type without behavior model is tied to a GPL behavior implementation - either via naming convention or explicit reference. Architecture models are parsed by MontiArcAutomaton, checked for well-formedness, and transformed into executable systems using generators for Java, Mona, and Python [10], [12].

## III. EXAMPLE AND PROBLEM STATEMENT

Reusing the commonalities of C&C software architectures for multiple similar systems facilitates efficient modeling. Consider two robots for exploration of unknown areas: one cheap and for indoor educational purposes, the other expensive and rugged for outdoor missions. Both feature different sets of sensors to detect obstacles, actuators to propel two parallel motors, and a navigation to control the robot based on the sensors' inputs. The platform-independent base software architecture for such a robot is depicted in Fig. 1. It comprises a composed component type `Explorer` that declares three subcomponents `col`, `dist`, and `ui` for sensors, a navigation controller `ctrl`, and two subcomponents `left` and `right` to access the parallel motors. The latter are of composed component type `ValidatedMotor` which itself declares two subcomponents `val` and `motor` to validate inputs and access motor drivers. The subcomponent declarations (SCDs) of `left` and `right` parametrize their respective `motor` SCDs with argument `100` as the component type `Motor` requires an integer as configuration parameter.

The behaviors of component types `Controller` and `Validator` are modeled platform-independently with automata. Depending on the actual platform properties, the GPL behavior implementations of component types `Color`, `Distance`, `HRI`, and `Motor` differ. Therefore, they are declared *abstract* which prevents ties to platform-specific GPL behavior implementations. Reusing this software architecture with both platforms demands for integration of proper behavior implementations for subcomponent declarations of abstract component types. To achieve this under reuse of the existing code generators the following is required:

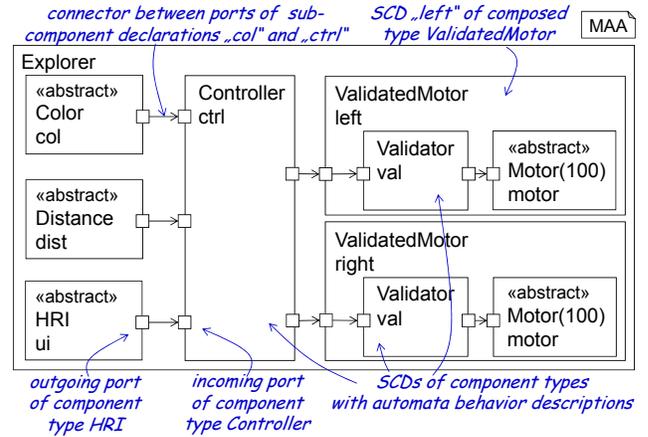

Fig. 1. C&C software architecture using abstract component types for different realizations of exploration robots. Port names and types are omitted for readability.

**R1** Additional parametrization: platform-specific components might require additional configuration, such as the hardware port a sensor is connected to. Introducing this information to the base software architecture would tie it to specific platforms again. Hence, it may not be defined within the platform-independent software architecture.

**R2** Behavior decomposition: Realizations of platform-specific components might be arbitrary complex and thus their decomposition is desired.

**R3** Architecture validity: The resulting platform-specific architecture must be a valid MontiArcAutomaton model, hence the platform-specific behavior implementations for abstract component types must be compatible to the abstract component types' interfaces.

**R4** Code generator compatibility: Retaining compatibility with existing code generators [10], requires integration to be performed completely prior to code generation and may not rely on generator specifics.

Exploiting the black-box nature of components to conceive subcomponent declarations of abstract component types as architecture extension points allows to fulfill these requirements with minor effort.

## IV. BINDING PLATFORM-INDEPENDENT COMPONENTS

Our approach allows the development of logical, platform-independent architectures and their transformation to platform-specific ones by binding abstract SCDs to platform-specific component types. To this effect, the architecture modeler describes extension points for different platforms by using abstract component types from respective *model libraries*. Afterwards, she selects or develops proper *implementation*

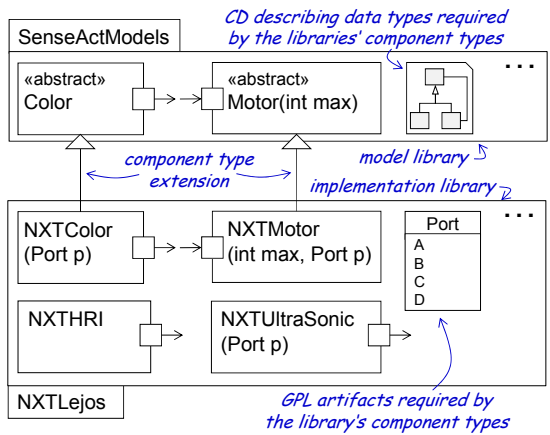

Fig. 2. Excerpt of the model library `SenseActModels` and the corresponding implementation library `NXTLejos` for NXT robots.

*libraries* that provide platform-specific realizations of the abstract component types. Modeling the *application configuration*, she defines how the SCDs of abstract types should be bound. Finally, MontiArcAutomaton processes the platform-independent software architecture, library components, application configuration model, and transforms the software architecture into a platform-specific model - without abstract components - according to the bindings. From this model, an executable system is generated.

Abstract component types are atomic and may not contain a behavior description, i.e., they are component interfaces with ports and configuration parameters. This follows the idea of abstract classes in object-oriented software engineering: they can be used during design time to describe properties expected from possible implementations, but they need to be extended and bound prior to code generation. To model a platform-independent software architecture, the abstract component types are imported from model libraries. Thus, a platform-independent software architecture may contain composed component types, atomic component types with behavior models, and abstract component types - all of which may use platform-independent data types only. Hence, the complete architecture is independent of GPLs and platforms.

Similarly to software architectures, model libraries may only contain composed component types, component types with behavior model, abstract component types, and data types. This ensures that model libraries are platform-independent and consequently that the importing software architectures remain platform-independent as well. Abstract component types of model libraries are realized via extension by platform-specific component types of implementation libraries, which may also contain platform-specific data types. Fig. 2 illustrates the relation between abstract and platform-specific component types in the context of their libraries: The model library `SenseActModels` contains abstract component types for sensors and actuators as well as class diagrams describing the required data types. The implementation library `NXTLejos` contains the platform-specific component types `NXTColor` and `NXTMotor`, which extend the abstract component types `Color` and `Motor`, respectively. Similarly, `NXTHRI` and `NXTUltraSonic` extend the component types `HRI` and `Distance` of Fig. 1 assumed in `SenseActModels`. The `NXTMotor` also introduces a new configuration parameter of type `Port` that describes the physical port the motor's hardware is connected to. This type is specific to the NXT platform and thus not part of the abstract `Motor` interface but provided by `NXTLejos` instead. Component types for different platforms might require other configuration and thus extend `Motor` differently.

Implementation libraries are referenced by bindings defined in *application configuration models* [13]. These models describe how abstract SCDs will be bound before code generation. Such models reference a single software architecture and contain a set of *bindings*. These map the architecture's abstract SCDs to platform-specific, parametrized component types, such that the bound component types inherit from the SCD's component type and that the arguments match the bound component type's parameters. Hence, platform-specific parameters are part of the bound component type and the application configuration, but not of the platform-independent software architecture.

```
                                        ApplicationConfiguration
1 import NXTLejosActuators.*;
2 application NXTExplorerApp for Explorer {
3   bind col to NXTColor(Port.A);
4   bind dist to NXTUltraSonic(Port.B);
5   bind ui to NXTHRI;
6   bind left.motor to NXTMotor(Port.C);
7   bind right.motor to NXTMotor(Port.D);
8 }
```

Listing 1. The application configuration `NXTExplorerApp` binds the abstract SCDs of architecture `Explorer` (Fig. 1) to platform-specific, parametrized types of `NXTLejos`.

Listing 1 illustrates the application configuration model `NXTExplorerApp`. It imports the implementation library `NXTLejos` (l. 1) before it declares its name and references the platform-independent software architecture `Explorer` (l. 2). Afterwards, it contains five bindings (l. 3-7) that describe how the abstract SCDs of `Explorer` should be replaced. Please note that the bindings for `left.motor` and `right.motor` (ll. 6-7) do not repeat the argument `100` passed to both `Motor` instances via their containing components (Fig. 1). Redefining arguments of the software architecture is prohibited and application configurations may define arguments for the platform-specific, bound component types only. Missing arguments are derived from the architecture and applied automatically.

With the libraries `SenseActModels` and `NXTLejos` and application configuration `NXTExplorerApp`, the platform-independent `Explorer` architecture can be transformed into the platform-specific software architecture depicted in Fig. 3. Here, the abstract component types used to describe the sensors and actuators have been bound to their platform-specific counterparts from the library `NXTLejos` and the arguments defined in the application configuration model have been applied. With different implementation libraries and additional bindings, the `Explorer` software architecture can be used with multiple target platforms. Mapping SCDs to component types and with platform-specific configuration parameters entails the following, updated, notion of bindings: a *binding* is a mapping from an abstract SCD to a parametrized, platform-specific component type such that this type and its parameters are applied to the SCD. As such, it consists of a source, which identifies a SCD in the architecture's hierarchy to be replaced, and of a target, which describes how it is to be replaced. The latter consists of a platform-specific component type and configuration arguments.

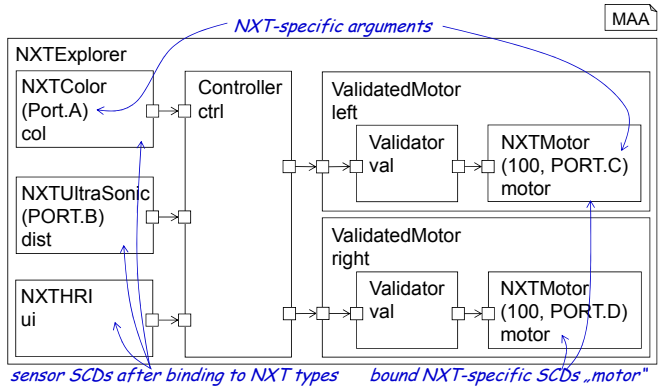

Fig. 3. NXT-specific architecture `NXTExplorer` with bound SCDs using platform-specific component arguments.

A binding for a MontiArcAutomaton software architecture $\mathcal{A}$ is a tuple $(s, T(a_0, \ldots, a_n))$, where:

- $s$ is a qualified name in $\mathcal{A}$ that identifies a subcomponent declaration of abstract component type $T_s$ with configuration parameters $p_0, \ldots, p_k$,
- $T$ is a platform-specific MontiArcAutomaton component type that inherits from $T_s$ and possibly adds configuration parameters $p_{k+1} \ldots, p_n$, and
- $a_0, \ldots, a_n$ is a list of configuration arguments, such that $a_i$ is of parameter type $p_i$.

Each element of $s = s_0 \ldots s_m$ refers to a unique SCD name starting from $\mathcal{A}$ (MontiArcAutomaton prohibits multiple SCDs of the same name in the same composed component [15]). Examples of valid names in the software architecture depicted in Fig. 1 are `col`, `left.val`, and `right.motor`. We write a binding $(s, T(a_0, \ldots, a_n))$ as s → T($a_0, \ldots, a_n$).

This notion of bindings enables to add platform-specific arguments to the resulting software architecture without tying the platform-independent base architecture to target platform properties (Req. R1). Furthermore, bindings may map abstract SCDs to composed component types. Hence, complex platform-specific behavior can be expressed by multiple interacting components (Req. R2).

Given the software architecture depicted in Fig. 1 and the libraries illustrated in Fig. 2, the bindings col → NXTColor(), left.motor → NXTMotor(10,Port.A), and right.motor → NXTMotor(10,Port.B) are valid bindings: the SCDs exist, the bound component types inherit from the SCDs abstract component types, and the arguments match. The following section describes how bindings are applied to a software architecture.

## V. BINDING TRANSFORMATION

Bindings are defined in application configuration models (cf. Lst. 1) that are processed by MontiArcAutomaton prior to code generation. These models are checked for well-formedness to ensure each bound SCD is abstract, bound exactly once, the component it is bound to extends the SCD's component type, and the passed arguments are valid. Nevertheless, bindings bind abstract SCDs – not component types – to platform-specific types and binding a SCD of a specific type differently is desirable and supported. Naively, this entails a component type with a single SCD of different component types – which conflicts with the notion of types in MontiArcAutomaton. Our binding transformation resolves these conflicts.

MontiArcAutomaton requires that SCD `motor` of component type `ValidatedMotor` has the same type in each instance of `ValidatedMotor`. Fig. 4 illustrates this with an excerpt of component type `Explorer` that shows the subcomponent declaration `left` and `right` of component type `ValidatedMotor` after applying the bindings left.motor → NXTMotor(10,Port.A) and right.motor → ROSMotor(10,Port.B), where `ROSMotor` is a component type applicable to be bound to `right.motor`. Afterwards, the component type `ValidatedMotor` is supposed to have a SCD `motor` of type `NXTMotor` (via `ValidatedMotor left`) and a SCD `motor` of type `ROSMotor` (via `ValidatedMotor right`). This naive transformation makes the type `ValidatedMotor` and with it the complete architecture invalid. We denote such type inconsistencies as *clashes*: There is a clash between two bindings $b_0 \ldots b_n \to T_b(a_{b_1}, \ldots, a_{b_x})$ and $c_0 \ldots c_m \to T_c(a_{c_1}, \ldots, a_{c_y})$ if they bind a SCD of a common parent component type to different component instantiations, i.e., SCDs $b_{n-1}$ and $c_{m-1}$ have the same type, $b_n$ and $c_m$ have the same name but $T_b \neq T_c$.

Desired bindings might clash and resolution prior to applying bindings is crucial to the resulting software architecture's validity. The following procedure takes care of clashes by replacing the types of all SCDs with new, unique types. To apply bindings, it conducts a breadth-first search through the component hierarchy defined by the root component type. During this search, the types and arguments of bound SCDs are replaced according to the bindings, i.e., bound. The types of unbound SCDs are replaced by copies of their original types with new and unique names to prohibit clashes. The corresponding procedure is depicted in Lst. 2.

Given a root component and a set of bindings, the procedure BIND visits all SCDs and either binds these according to the bindings or replaces their type with a new, unique type based on the original one. To this effect, the procedure utilizes a stack of tuples of names and component types. Initially the stack contains only the empty qualified name and a copy of the architecture's root component (ll. 2-4). The copy's type name is ensured to be unique by function `uniqueCopy()`. Afterwards it iterates over the stack's tuples and (ll. 5-14) inspects every SCD of the currently visited component type (such as `ValidatedMotor`). The qualified name `q` is updated with the current prefix and concatenated with the actual SCD's name using a ternary operator (l. 8, for instance to `left.val`) and it is checked whether a binding for the SCD indicated by `p` exists (l. 9). If a binding exists, the type and the arguments of the actual SCD are changed accordingly (ll. 10-11). As the replaced SCD's type must be abstract (and hence atomic) and the replacing component type must be platform-specific (it may be composed but not contain abstract SCDs), visiting the bound new component type is not necessary. In case there is no binding for the actual SCD, its type is set to a unique (in terms of its name) copy of itself (ll. 13-14). Finally, the currently updated hierarchy, as defined by `newRoot`, is returned (l. 15) for further analyses and code generation.

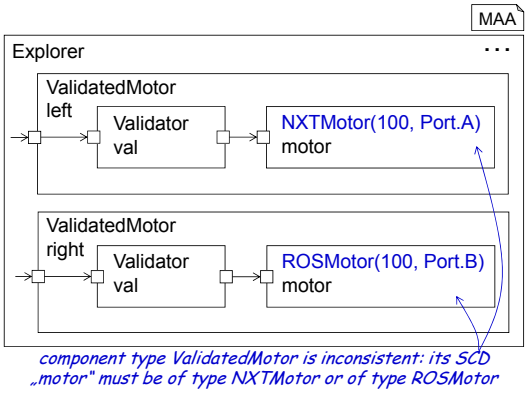

Fig. 4. Example for a clash between the two bindings `left.motor` → `NXTMotor(10,Port.A)` and `right.motor` → `ROSMotor(10,Port.B)`, which our transformation resolves.

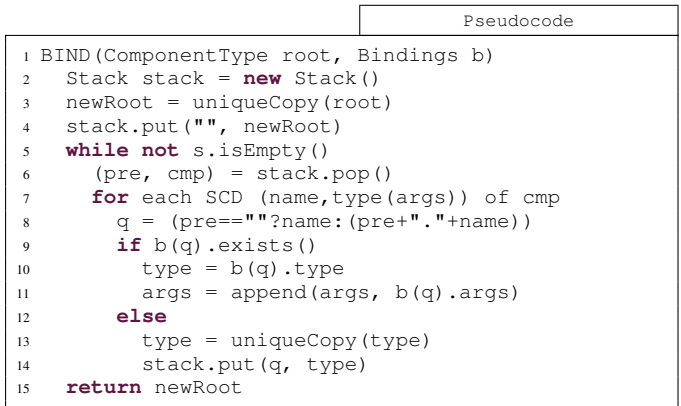

Listing 2. The procedure `BIND` replaces the types of all SCDs with either bound types or new, unambiguous types.

This procedure can be performed prior to any code generation and returns a valid MontiArcAutomaton software architecture (Req. R3) that describes the platform-specific architecture completely. Hence, the architecture can be processed by existing code generators without need for modifications (Req. R4). The procedure prohibits clashes but produces new component type definitions (l. 3 and l. 13) for each non-abstract subcomponent declaration. The number of new component types is thus bound by the number of subcomponent declarations. Whether this influences the number of artifacts in the generated system however depends on the employed code generators and their translation from component types to artifacts.

## VI. RELATED WORK

The presented approach is related to our previously introduced approach, deployment modeling, and other ADLs.

Our previous approach [13] relied on exchanging behavior implementation GPL artifacts instead of component types. Consequently, it could not produce software architecture models employing with different platform-specific component types. The architectures' components referenced to different behavior GPL artifacts instead. This prohibited to introduce new arguments to SCDs. Exploiting the notion of component inheritance lifts bindings completely to model level and enables such arguments while retaining a type-safe architecture. Handling references to different behavior GPL artifacts is no concern for code generators anymore and with code libraries, the library property models of [13] have become obsolete as well. These models described which abstract component types the contained behavior implementations belong to and identified the required run-time system (the GPL machinery required to enable system execution [12]). Now both is made explicit in the component types via inheritance and a new component property. Hence, libraries can also contain platform-specific component types for multiple run-time systems.

Bindings are related to deployment of C&C architectures to specific platforms [9], [16], but differ in the level of abstraction: deployment maps components to elements of the participating platforms and thus requires explicit platform models. Additionally, deployment may consider proper code generation for specific target platform elements, proper realization of connectors between physically distributed components, or mechanical and electrical properties of the target platforms. This imposes platform expertise on the application modeler.

The xADL [17] encourages including implementation details in component models. While omitting this allows describing platform-independent architectures, we are unaware of any similar pre-generation transformation. Relations to other ADLs and "abstract platforms" of MDA [7] are discussed in [13].

## VII. DISCUSSION

Application configuration models specify single bindings per SCD. For large architecture this is inconvenient, but can easily be solved by binding abstract component types and calculating the actually affected SCDs. This however is only part of improving the application configuration modeling language: additional features under consideration are conditional expressions over architecture properties and rewiring connectors for multiple interconnected bound component types. Also, interfaces of abstract component types need to be broad enough to support arbitrary platform-specific component types. They are by design, as the software architecture defines what is required. Furthermore, we do not bind non-abstract component types. While possible with this approach, this allows changing the architecture beyond recognition. This is not yet intended. Furthermore, we currently do not allow to bind SCDs of composed component types. While interesting, this leads to issues for abstract composed component types that contain abstract component types. The procedure BIND retains the processed software architecture's validity by introducing new component types to avoid clashes. Consequently, the resulting architecture contains redundant component types. We currently investigate a less invasive procedure that iteratively detects clashes and solves these introducing new components types only where necessary.

## VIII. CONCLUSION

We have presented an enhanced approach to transform platform-independent into platform-specific software architectures. This approach builds upon previous work [13] and lifts it to model level completely. It applies bindings from abstract SCDs to parametrized, platform-specific component types of a software architecture and produces a valid software architecture again. The presented procedure is type-safe, allows to incorporate platform-specific configuration, reduces the complexity of MontiArcAutomaton code generators, and enforces a strict separation between platform-independent and platform-specific constituents. We are currently investigating the expressiveness of the new approach in further case studies.